\documentclass[aps,twocolumn,amsmath,prx,showkeys]{revtex4-1}
\usepackage{graphicx}
\usepackage{subfigure}
\usepackage{amsmath}
\usepackage{bibentry}

\begin{document}

\title{An intrinsic topological model in the absence of symmetry}

\author{Ye Xiong} 
\email{xiongye@njnu.edu.cn}
\affiliation{Department of Physics and Institute of Theoretical Physics
  , Nanjing Normal University, Nanjing 210023,
P. R. China}

\begin{abstract}

	We study the vibrational spectrum of a constrained classical ring. Due to the
	presence of 2-order exceptional points, a topologically trivial band at the
	infinity can make the vibrational band topologically nontrivial.  The symmetry,
	which is believed to be indispensable in topological models, is absent in this
	model.  The fractional boundary states can be found in such classical system.
	Furthermore, the other aspect of the bulk boundary correspondence is revealed: an
	extra fractional exceptional point is topologically protected in bulk by the
	boundary states.

\end{abstract}

\keywords{non-hermitian, exceptional point, fractional state}
\maketitle

\section{Introduction} 

The exceptional point (EP) is playing very important rules in the non-hermitian
Hamiltonian systems\cite{Moi, Berry2004, Heiss2012, Mehri-Dehnavi2008, Liang2013,
Malzard2015, Cerjan2016, Lin2016a, Xu2016e, RevModPhys.93.015005}. It can be obtained by
further splitting a 2-fold degenerate weyl point when the Hamiltonian becomes
non-hermitian.  The bands in the region between these EPs are connected so that one can
reach the other bands smoothly by varying the wave vector $k$ {\it around} the EP
continuously\cite{RevModPhys.93.015005}. Although the exceptional ring, a ring of EPs in
higher dimensional Brillouin zone (BZ), has been
discussed\cite{PhysRevLett.118.045701,Cerjan2019}, most of the physicists avoid to perform
the arguments right at the EP because the Hamiltonian at that point is defective and
cannot be diagonalized into the distinct bands completely. As a result, the project
operator on the bands and the metric matrix in the eigenstate representation are both not
well defined at the EP.  But this is not the case when the infinity emerges. Because a
meaningful state cannot live in the subspace with the infinite eigenvalue, one can avoid
to define the metric matrix there. Now if the eigenvalue of the Jordan matrix (the part of
the Hamiltonian that cannot be diagonalized) is approaching to the infinity, it is
possible to redefine the metric matrix to measure the length of the states only in the
remaining subspace with the finite eigenvalues.

\begin{figure}[ht] 
  \includegraphics[width=0.39\textwidth]{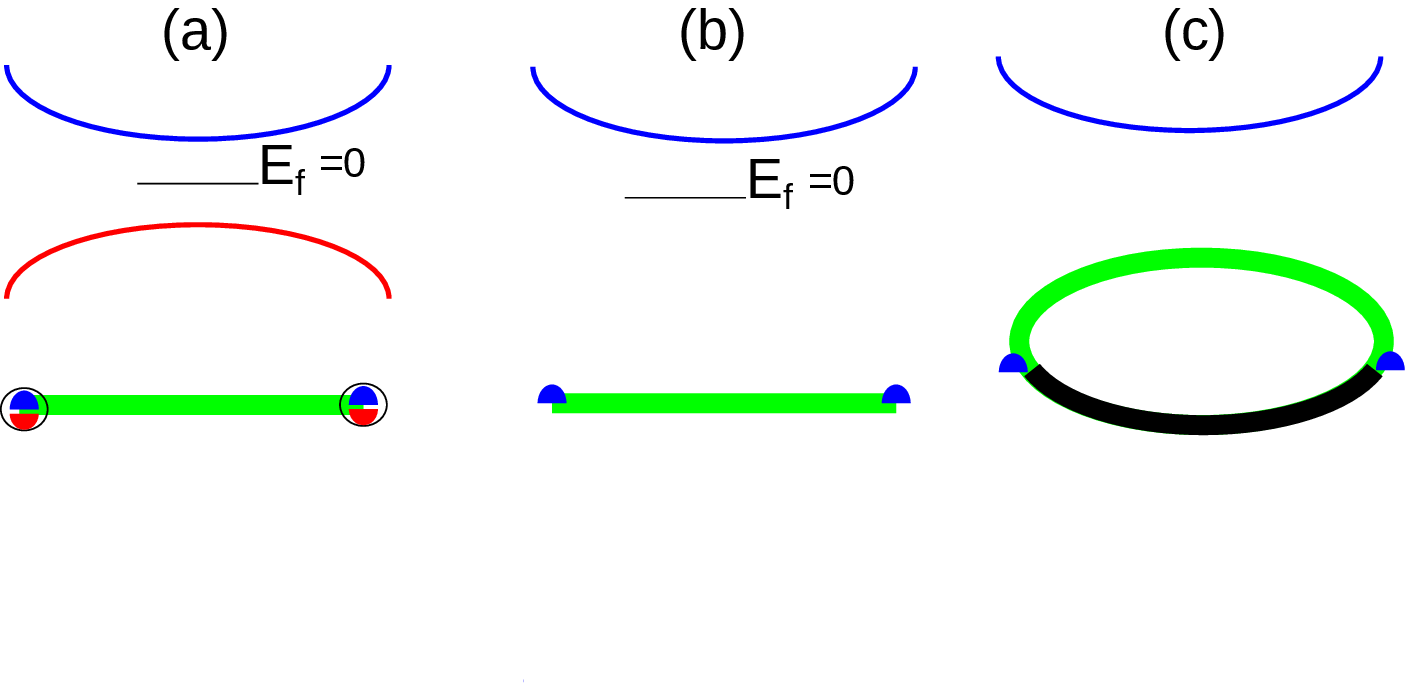}
  \caption[Fig]{\label{fig1} (a) In the upper panel, the two bands in the $k$ space for
	  the topological SSH model are shown. In the lower panel, the chain in the open
	  boundary condition has one boundary state at each end, which is composed by the
	  two half states (colored half circles) contributed from the two bands,
	  respectively. (b) In the upper panel, the band for the 1D topological
	  superconducting model is shown. Its duplication in the Nambu representation
	  below the fermi energy is not plotted.  In the lower panel, the  fractional
	  modes (majorana fermions) at the ends will couple nonlocally to form one
	  majorana zero mode. (c) In the upper panel, the vibrational band of our
	  classical model is schematically shown. There is only one band at the finite
	  value. The shape of the line has nothing to do
	  with the actual dispersion of the band and is only used to indicate the
	  nontrivial zak phase of the band. The non-hermitian skin effect will destroy the
	  topology of the bulk states in the open boundary condition. So in the lower panel,
	  we consider a  ring with one part in the topologically trivial phase. Here the
	  green part is the chain of our model and the black part is a traditional elastic
	  string in the topologically trivial phase. The fractional modes should appear at the 
  two interfaces.}
\end{figure} 

In the most cases, the states whose eigenvalues are at the infinity will have nothing to do with the
remaining states whose spectrum are at the finite values (SAFV). The Hamiltonian can be
written as
\begin{equation}
	H=\begin{pmatrix}
		H_1 & V \\
		U & H_2
	\end{pmatrix},
	\label{H1}
\end{equation}
where $H_1$ is the Hamiltonian in the subspace with the finite spectrum, $H_2$ is that
with the infinite spectrum and their couplings are $U$ and $V$, respectively. After
projecting out the subspace of $H_2$, the finite eigenvalue $\omega$ of $H$ can be
obtained by diagonalizing the matrix $H_1+ V(\omega -H_2)^{-1} U$.  As $H_2$ is
approaching to the infinite matrix, $(\omega -H_2)^{-1}$ is the zero matrix so that the
second term in the above expression is zero. It is the $H_1$ itself that provides the
SAFV and the corresponding eigenstates of the Hamiltonian. But when $H_2$ is defective,
e.g., $H_2 = \infty^2 \begin{pmatrix} 0^1 & 1 \\ 0 & 0^1 \end{pmatrix}$, then $(\omega
-H_2)^{-1} = \begin{pmatrix} -0^1 & 1 \\ 0 & -0^1 \end{pmatrix}$, where the superscript
indicates the order of the infinity and the infinitesimals with the rule $0^1 \infty^1
=1$. So if the EP appears in $H_2$, the second term is not zero matrix any more and the
states at the infinity must be considered in the diagonalization. As a demo model, we will
consider the vibrational spectrum of a classical chain. At any $k$ in the BZ, the
equations of motion (EOM) have $3$ linear equations for the $3$ unknown variables.  One of
the eigen-frequencies is at the finite value and the other two are at the infinity with only one
left (right) eigenstate. So there is an exceptional BZ: a 2-order EP at every point in
the BZ.  The single band at the finite value is topologically nontrivial, a band with
$\pi$ zak phase. We want to emphasize that the zak phase of the infinite band is $0$ so
that even counting the band at the infinity, the total zak phase is also in a nontrivial
value. In this sense, the topology in this model is intrinsic: symmetry is not needed to
balance a pair of bands to protect the nontrivial topological index. 

There is also correspondence between the topology of the bulk states and the boundary
states at the ends. In Fig.  \ref{fig1}, we begin to present our arguments from the other
well-known 1-dimensional (1D) models. The spectrum of the 1D Su-Schrieffer-Heeger (SSH)
model\cite{Su1979, Su1980, Kivelson1982, Li2014a, Rhim2017} are illustrated in the upper
panel in Fig. \ref{fig1} (a). The two bands are denoted by the red and the blue lines,
respectively. The zak phase of each band is $\pi$ in the topologically nontrivial phase.
The chiral symmetry is needed to balance the two bands. When the open boundary condition
is taken (shown in the lower panel), each band leaves one half state at the end which is
denoted by the colored half circle in the figure. The topologically protected boundary
states (large circles) are the combination of the two half states at each end. In Fig.
\ref{fig1} (b), we show the band of a 1D topological superconducting chain. There is only
one band because we have omitted its duplication below the Fermi energy $E_f=0$ in the
Nambu representation.  The chiral symmetry is present implicitly between the band and its
duplication. When the boundaries are introduced, the single band leaves {\it one} half
state at each end.  These fractional states are the majorana states and their nonlocal
combination is the majorana zero mode. 

The vibrational band of our classical model is schematically shown in the upper panel in
Fig.  \ref{fig1} (c). The exact vibrational spectrum are on a complex plane and the shape
of the line here has nothing to do with the dispersion of the spectrum. It is only used to
indicate that the zak phase of such band is $\pi$. The chiral symmetry is absent because
there is only one band at the finite value. Adopting the similar arguments in the previous
subfigures, it seems that we can find isolated fractional states at the ends of the chain.
But, unfortunately, the non-hermitian skin effect emerges first and destroys the topology
of the bulk
states\cite{Silveirinha2019,PhysRevB.99.201103,PhysRevB.102.085151,Helbig2020,Zhang2021}.
To avoid the skin effect, we embed a topologically trivial chain (black part of the ring
in the lower panel) in our chain. The interfaces between the regions with different
topological phases are equivalent to the open ends in the topological sense. But due to
the absence of the chiral symmetry, we don't find a zero frequency mode. Instead, the
whole spectrum of the black part of the ring are modified.  Such phenomenon breaks some of
our knowledges on the 1D vibrational chain. The black part of the ring is in the
topologically trivial phase and is equivalent to a normal elastic string in our
traditional life. The textbooks on phonons tell us that the local decorations, including
the connections to the other chains at the ends, can only introduce a few local
vibrational modes. For what reason the connections to the green chain at the two far ends
can modify the bulk spectrum of an elastic string? 

We find that it can be traced back to the fractional states at the interfaces and the
origin of the topology in the green part. First, as that has been discussed in the
previous models, the nontrivial zak phase in the green chain ensures that there are
fractional states at the interfaces.  Second, the bulk boundary correspondence actually
declares that the topologically protected boundary states are caused by a kind of split of
the states in the bulk topological band. While in our model, the bulk topological band is
related to the band at the infinity through the help of EPs. The similar relation should also
exist behind the fractional boundary states: one more state at the infinity is required to
accompany with the boundary states.  For a ring with $L_1$ unit cells in the green part, we
find $L_1-1$ EPs (remember that there is one  2-order EP at each $k$ which will map to
$L_1$ EPs in the real space representation) are still kept in the 2-order.  But one of the
EPs is changed from 2 to 3 order. This is topologically equivalent to an appearance of a half
2-order EP. Such phenomenon is universal, regardless of the details near the interfaces,
the parameters of the model and the disorders in the bulk. So this model actually reveals
a boundary bulk correspondence: an extra fractional EP in bulk is
topologically protected by the boundary states. This explains the spectrum modification
because the whole spectrum must be modified dramatically when a new EP appears\cite{xiong1}.
The eigenstates of these modified spectrum are spatially localized at one interface. But
they can respond to the global effective magnetic flux so that the nonlocal correlation in
between the fractional states is still present in such classical system.  

\section{ a 1D single band topological lattice}

\begin{figure}[ht] 
  \includegraphics[width=0.46\textwidth]{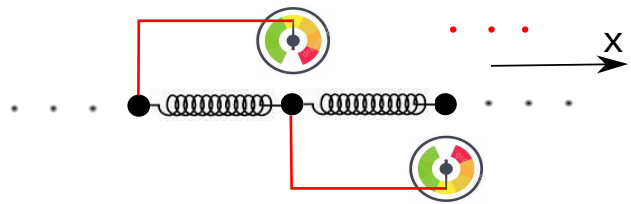}
  \caption[Fig]{\label{fig2} (a) Our model is based on a 1D classical vibrational chain.
	  The nearest neighboring MPs are connected by the springs. On the top of each MP,
	  there is a speedometer that measuring its velocity. The rigid bonds
	  are connecting each MP with the pointer of the speedometer in its front. This
	  will  restrict the displacement of the $i$th MP, $x_i$, with the velocity of the
  $i+1$th MP, $\dot{x}_{i+1}$, by the constraint $x_i - \dot{x}_{i+1} =0$.}
\end{figure} 

\begin{figure}[ht] 
  \includegraphics[width=0.36\textwidth]{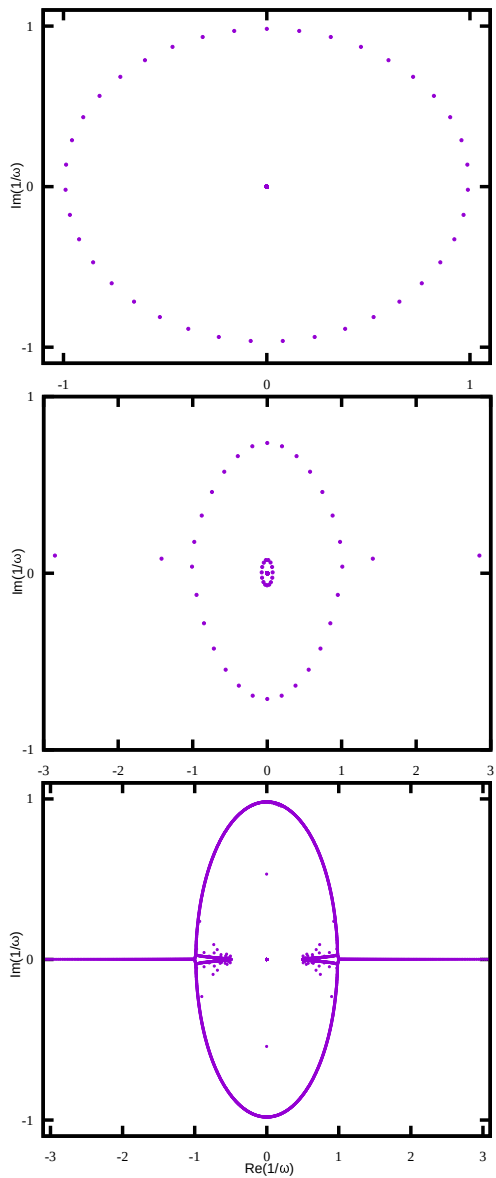}
  \caption[Fig]{\label{fig3} The inverse spectrum $\frac{1}{\omega}$ for the ring shown in
	  the lower panel in Fig. \ref{fig1} (c). The lengths of the topologically
	  nontrivial (green) part and topological trivial (black) part are $L_1$ and
	  $L_2$, respectively. (a)
	  $L_1=39$ and $L_2=1$. (b)
  $L_1=32$ and $L_2=8$. (c) $L_1=2000$
  and $L_2=1000$. The inverse spectrum on the real axis outside $[-3:3]$ are not shown.}
\end{figure}

The 1D classical lattice is made up by the identical mass points (MPs) with
the mass $m=1$.  Each MP can only move in the direction along the chain in the $x$
direction. The springs, whose coefficients are $K=1$, are connecting the nearest neighboring
MPs.  The active constraints (ACs), $x_i-\dot{x}_{i+1}=0$, are restricting the
displacement of the $i$th MP with the velocity of the next one. They are realized by the
rigid bonds between the $i$th MP and the pointer of the speedometer on the top of the
$i+1$th MP, as the Fig. \ref{fig2} shows.  For each MP, there are $3$ unknown variables:
its displacement $x_i$, its velocity $\dot{x}_{i}$ and the stress force in the bond
$\lambda_i$. The linear first order time derivative EOM for the $i$th MP are
\begin{eqnarray}
	\dot{x}_i & \equiv & \dot{x}_i \\
m\ddot{x}_i \equiv m \dot{\dot{x}}_i & = & K(x_{i+1}+x_{i}-2x_i)
	+\lambda_i \\
	x_i -\dot{x}_{i+1} & = & 0.
	\label{eq1}
\end{eqnarray}

We want to emphasize that one will find a wrong EOM by applying the Lagrange multiplier
directly in the Lagrangian.  This is because the Newton's third law is not obeyed on the
chain itself: the reaction force in the bond does not act back on the $(i+1)$th MP
but on the pointer of the speedometer. After the Fourier transformation on the $3$
variables,  we can find the EOM for $|\psi_k \rangle \equiv (x_k, \dot{x}_k,\lambda_k)^T$,
\begin{equation} 
	\begin{pmatrix} 1 & 0 & 0 \\ 0 & m &0 \\ 0 & 0 & 0 \end{pmatrix}
\dot{|\psi_k \rangle} = \begin{pmatrix} 0&1&0 \\ -4K\sin^2(k/2)& 0 & 1 \\ 1&
-e^{ik}&0 \end{pmatrix} |\psi_k \rangle.  \label{eom} 
\end{equation}

After substituting $|\psi_k \rangle = e^{-i\omega t} |\psi_k(\omega) \rangle$, the
eigen-frequency $\omega$ and the eigenstate $|\psi_k(\omega) \rangle$ are found by
solving a general eigen-problem likes, $ -i\omega C |\psi(\omega)\rangle = D
|\psi(\omega) \rangle$, where $C$($D$) is the $3\times 3$ matrix on the left(right) side of Eq.
\ref{eom}. Here $C$ is an irreversible matrix and $D$ is reversible. By multiplying
$i\frac{D^{-1}}{\omega}$ from the left, the above equation is equivalent to the
eigen-problem of a non-hermitian matrix $D^{-1}C$, 
\begin{equation}
\frac{i}{\omega} |\psi_k(\omega)
\rangle = D^{-1}C | \psi_k (\omega) \rangle.
\label{noh}
\end{equation}

We find that the matrix $D^{-1}C$ is defective and its eigenvalues are, $e^{ik}$ and
$0$.  The first one describes the vibration band whose dispersion is $\omega= i e^{-ik}$.
The other is the band at the infinity with a 2-order exceptional point. The eigenstates
for the SAFV are $ |\psi^1_k \rangle =( 1, e^{-ik}, m e^{-2ik}+4 K\sin^2(k/2))^T$, while
that for the band at the infinity is $( 0, 0, 1 )^T$. This result is what we expected in
the introduction: there is an EP at every $k$ in the BZ and the corresponding spectrum are
at the infinity. 

In the presence of EP, one will fail to decompose an arbitrary state into the
superposition of the eigenstates. Just as the above equations show, the dimension of the
Hilbert space is $3$ at every $k$ while there are only $2$ available (right) eigenstates.
But our model is special at the point that the spectrum corresponding to the EP are at the
infinity. All modes carrying the infinite vibrational frequency are unobservable in the
real experiment. The available state in the Hilbert space is just $c |\psi^1_k\rangle$,
where $c$ is a real number. So the effective dimension of the Hilbert space is one and
this gives us more freedom to define the metric matrix $\hat g$, which measures the length
of a vector in the Hilbert space by $ \langle \psi | \hat g | \psi \rangle$. Actually, any
hermitian matrix with nonzero $ \langle \psi^1_k | \hat g | \psi^1_k \rangle$ is a good
candidate. But because we are going to connect the chain to a traditional elastic string,
to avoid further discussions on the domain wall between regions with different metric
matrix, the metric matrix here is taken as 
\begin{equation}
	\hat g = \frac{1}{2} \begin{pmatrix} 1 & 0 & 0 \\ 0 & 1 &0 \\ 0 & 0 & 0 \end{pmatrix},
	\label{metric}
\end{equation}
will be the same as that in a topologically trivial chain.

The berry phase is defined as $-i \langle \psi^1_k | \hat g \frac{\partial}{\partial k} |
\psi^1_k \rangle$ and the zak phase is the integration of the berry phase in the BZ: 
\begin{equation}
	\int_0^{2\pi} -i \langle \psi^1_k | \hat g \frac{\partial}{\partial k} |
\psi^1_k \rangle dk. 
	\label{zak}
\end{equation}
It is easy to verify that this zak phase is $\pi$, at the topologically nontrivial value. 

By changing the representation for $k$ space to the real space, for a ring with $N$ unit
cells, there are $N$ inverse eigen-frequencies on the unit circle on the complex plane.
They are belong to the band at the finite value.  Simultaneously, there are also $2N$
inverse eigen-frequencies at $0$ that corresponding to the $N$ 2-order EPs with the
spectrum at the infinity. When the open boundary condition is taken, a non-hermitian skin
effect appears and destroys all bulk states of the ring. 

To avoid the skin effect, we embed a topologically trivial chain into the
ring. It is realized by erasing a sequence of ACs in our model because in the absence
of AC, the chain is just the simplest model on phonon in many textbooks. It is
absolutely topologically trivial and can be considered as the simplified model of a
traditional elastic string. 

We first show the spectrum for the ring with $L_1=39$ and $L_2=1$ in Fig. \ref{fig3} (a).
It is realized by erasing one AC on the perfect ring.  In most of our beliefs, when a
constraint is erased, one more degree of freedom is left free so that a new vibrational
mode with the finite frequency should be observed on the spectrum. But the result is
surprisingly opposite. In the figure, there is $39$ points near the unit circle.  They are
the bulk states with the SAFV in the green part. Besides these, the rest states are at
$\frac{1}{\omega}=0$.  This means, when one AC is removed from the ring, one more
vibrational mode is frozen, instead of being released. 

When we increase the length of black part to $L_2=8$, as the Fig. \ref{fig3} (b) shows,
the vibrational modes are not frozen exactly. But their frequencies are still strongly
amplified. The inverse spectrum of the green part are on a circle a little smaller than the unit
circle. This is the finite size effect. They will approach to the unit circle as
increasing $L_1$.

In Fig. \ref{fig3} (c), we show the case when $L_1=2000$ and $L_2=1000$. For a long
isolated black chain, its spectrum are $\omega = \pm 2\sqrt{\frac{K}{m}} \sin(k/2)$ in the $k$
space. So besides the boundary states, the bulk spectrum should be in the range $[-2,2]$ on
the real axis. But our results show that all the inverse spectrum within the unit circle
have been altered from the real axis. The percentage of these modified states is kept at about
$66\%$ and does not decrease as increasing $L_2$. 

Now we summarize the common features in the above subfigures. For a ring with $L_1$
topologically nontrivial part and $L_2$ topologically trivial part, there are $L_1$ states
with the inverse spectrum on the unit circle. This means that the bulk states in the
green part are not modified by the embedment. It will ensure the validity
of the topological arguments in the following. 

As we have discussed in the introduction, the existence of the fractional states at the
interfaces is protected by the nontrivial topology in the green bulk. These fractional
states may couple together nonlocally to form one state, just like that in the topological
superconducting case. Although there is no chiral symmetry to fix its energy to zero, the
above inverse spectrum still confirm its existence. The trace of the $D^{-1}C$ matrix,
$Tr(D^{-1}C)$, is a real number.  This means the summation of the spectrum is an imaginary
number. In most cases, the eigenvalues will appear in pair with $\pm
Re(\frac{1}{\omega})$. If we look at the imaginary axis with $Re(\frac{1}{\omega}) =0$,
besides the spectrum at the infinity that corresponding to the EPs, there are even number
of eigenstates on the axis. But if further eliminating the bulk states in the
green part, there is only {\it one} state left on the imaginary axis. This confirms our
conclusions on the nontrivial topology in the green bulk and the fractional states at the
interfaces.

There are also $L_1$ EPs accompanied with the above bulk states. Their spectrum are still
at the infinity, $\frac{1}{\omega} =0$.  But different from that in the $k$ space, only
$L_1-1$ of them are still the $2$-order EPs. The rest one EP is changed to the $3$-order
EPs. This feature is robust against the disorders in the ring or the details near the
interfaces.  So mostly, such feature should also have a topological explanation.  When the
boundary condition of a topological model is changes from open to periodic, the extra
hoppings between the boundaries will coupling the boundary states back to the
topologically nontrivial band states. In another word, the boundary states are a kind of
split of the band states, i.e., the edge states in 2D quantum Hall system can be
considered as the split of bulk state with different chirality or different sign of
velocity and the end state of the SSH model is half from the upper band and half from the
lower band. So it seems that the two fractional boundary states in our model are a split
of one bulk state in the topological band and they must also inherit one 2-order EP at the
infinity because every state in the band is accompanied with one EP. But different from
the case of the SSH model, as Fig. \ref{fig3} (a) shows, a hopping between the fractional
states, which is realized by the spring here, cannot recombine them. The bulk state can
only be restored when the missing constraint is added back.  But one more variable, the
force in the bond, is enrolled in this process. So the boundary states should only
accompany with one dimensional subspace with infinite eigenvalue instead of two
dimensional. To keep the connection between the states at the infinity and the finite
value, such one dimensional subspace should also be the eigen-space of an EP. So the
fractional boundary states are accompanied with a {\it half} $2$-order EP.  This is
consistent with our numerical results because a $3$-order EP is equivalent to
$1+\frac{1}{2}$ $2$-order EP topologically. So besides the bulk states that can protect
the existence of the boundary states, the boundary states can also protect an extra state
in the infinite bulk band. 

\begin{figure}[ht] 
  \includegraphics[width=0.36\textwidth]{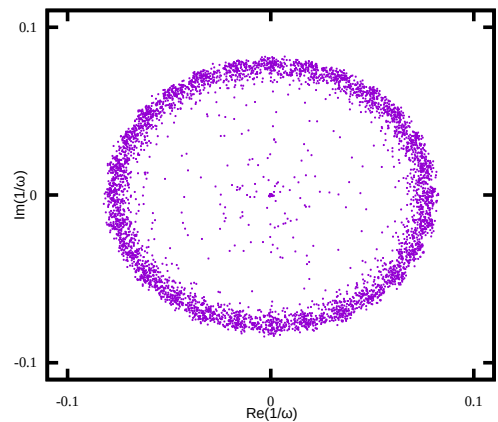}
  \caption[fig]{\label{fig4} (a) the evolution of inverse spectrum with $k'$ for the ring with
	  $L_1=32$ and $L_2=8$. Those spectrum on the unit circle are varying on the circle
	  smoothly and are outside the visitable region of this figure. The modified
	  spectrum do respond to the effective magnetic flux and they are varying with
  $k'$ chaosly.}
\end{figure}

The appearance of a new EP or the modification of the order of an EP must be accompanied with the
deformation of spectrum on the complex plane\cite{xiong1}. This is why the spectrum of the
black part is different from that of a free chain. By focusing on the eigenstates of those
modified spectrum, we find that they are localized spatially at one interface, instead of
extended in the black bulk or localized at the two interfaces simultaneously. But they
still represent the nonlocal correlation between the fractional states in between the two
interfaces because they can respond to the effective magnetic flux $e^{ik'}$ in the ring.
At the first sight, a magnetic flux is inapplicable in a classical vibrational ring. But
if we consider the ring in Fig. \ref{fig1}(c) as one supercell and build a long lattice
with this supercell, $k'$ here is the wave vector for this lattice and the effective
magnetic flux $e^{ik'}$ is added naturally during the calculation of  the spectrum in the
$k'$ space. The inverse spectrum are plotted in Fig. \ref{fig4} for $L_1=32$ and $L_2=8$.
As $k'$ is varying from $0$ to $2\pi$, the spectrum of the green part are varying smoothly
on the unit circle. It further confirms that the spectrum on the unit circle are
correspond to the bulk states in the topologically nontrivial region. But the inverse
spectrum inside the unit circle, whose eigenstates are localized at one interface, are
still changing with $k'$. Such behavior should not exist in the absence of extra
fractional EP because the effective magnetic flux can be gauge transfered to a phase
jump at the sites far away from the localized states. The localized states cannot feel
this phase jump so that they cannot respond to the effective magnetic flux.  But when
the EP is present, the spectrum can be very sensitive to the parameters so that the
localized states can feel the phase jumping far away from them. This also explains why
these inverse spectrum are varying chaosly with $k'$.

\begin{figure}[ht] 
  \includegraphics[width=0.36\textwidth]{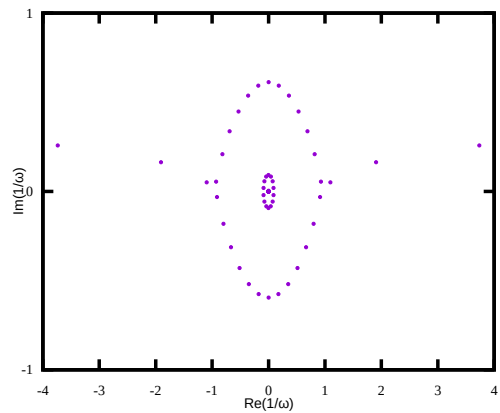}
  \caption[fig]{\label{fig5} (a) the inverse spectrum when both the spring constant and
	  the constraint are randomly distributed along the ring. The length of the ring is
  the same as that in Fig. \ref{fig3} (b).} 
\end{figure}

All the above conclusions are robust against disorder. We consider both constraint
disorder and spring disorder: the constraint at the $i$th site is $x_i - w_i \dot{x}_{i+1}
=0 $ and the spring constant between the $i$th and the $i+1$th sites is $K_i = z_i K$,
where $w_i$ and $z_i$ are randomly distributed in $[0,2]$ and $[0.5,1.5]$, respectively.
We find the inverse spectrum in Fig. \ref{fig5} are qualitatively same as that in Fig.
\ref{fig3} (b). Here all other parameters are the same in these two figures.

\section{Conclusions and discussions}

We find a new class of topological models that are not protected by the symmetry.  In
traditional topological models, including non-hermitian models, one needs to split the
bands into two groups. In the fermion system, the bands are naturally separated into the
valence bands and the conduction bands by the fermi energy. In the boson case, one still
need to specify the gap that separates the two groups of bands. This gap or the fermi
energy is never taken on the top or the bottom of the whole bands because the project
operator on one group of the bands becomes $\hat P =1$ or $\hat P =0$ in this case. Such
group of the bands must be topologically trivial because its project operator commutes
with the translation operator in the reciprocal space, $e^{i\hat r \cdot \delta k}$, where
$\hat r$ is the position operator and $\delta k$ is an infinitesimal displacement along
the Wilson loops in the BZ\cite{Yu2011a, Soluyanov2011, Kivelson1982a, Zhao2014,
Marzari1997, Taherinejad2014, Li2015k, Gresch2016, Liu2016q, Alexandradinata2016,
Lian2017}. So one must split the bands in the middle gap, and the symmetry is needed to
balance these bands intergrouply or intragrouply. But in our model, there is only one band
at the finite value. The split of bands into two groups is not needed and not even
applicable. As a result, symmetry is not the key ingredient in this class of topological
model. 

To be topologically nontrivial, the project operator on this single band should not be
$\hat P =1$. This can be realized only when there is an invisible band at the far infinity
and there are EPs to effectively couple these two bands. In principle, the EPs with SAFV
may also play a similar rule. But one cannot define the length of the eigenstates in the
full Hilbert space in that case.

The model shows many interesting behaviors: it shows fractional states at the interfaces
in a classical chain, it reveals another aspect of the bulk boundary
correspondence that the topologically protected boundary states further protect the
existence of the fractional EP in bulk, the spectrum are varying with the effective
magnetic flux chaosly in such linear system. Besides these, the model also leaves us with
many interesting questions. For instance, due to the freedom in the definition of metric
matrix, it is possible to find a topologically protected boundary state on an interface in
between two topologically trivial regions. Such boundary state should be traced back to
the metric matrix interface instead of the interface for the topology of the bulk states.
The model can also be  extended to the higher dimension. We have found that the in-gap
boundary states are protected to be extended in bulk on a 2D lattice with nonzero Chern
number.

One of the weak points of the present model is the unstableness: there are spectrum on the
upper half plane of the complex plane. So one can only observe the phenomena in a short
time before the chain falls apart in experiment. We are looking for a way to stabilize the
chain. If we succeed in this, such model will have potential application in engineering:
one can obtain a stiff chain (with higher vibrational eigen-frequencies) from a soft chain
by connecting the two ends to our model.

{Acknowledgments.---} 
The work was supported by the 
National Foundation of Natural Science in China Grant Nos. 10704040.

\bibliographystyle{apsrev4-1}
\bibliography{main}

\end{document}